# Missed 2020 Superoutburst of the Eclipsing WD-BD Cataclysmic Variable SDSS J1433+1011

**D. V. Denisenko**

Sternberg Astronomical Institute, Lomonosov Moscow State University, Russia;
Vorobyovy Gory Education Center, Moscow, Russia;
e-mail: d.v.denisenko@gmail.com

Cataclysmic variable SDSS J143317.78+101123.3 with $P=0^d.054241$ (78.1-min) was suspected to be a possible dwarf nova of WZ Sge type with the sub-stellar donor, but without detected outbursts so far. Checking the newly available data from ATLAS survey has revealed the outburst by at least 6 magnitudes in September 2020, thus confirming the dwarf nova nature of this object with the brown dwarf secondary. Other projects and individual observers have stopped their monitoring of this target several days before the outburst. This finding strengthens the value of observing the twilight zone by the professional surveys and amateurs.

## 1  Discovery and previous study

Cataclysmic variable SDSS J143317.78+101123.3 (J1433+1011 for short) was discovered by the Sloan Digital Sky Survey (Szkody et al., 2007) based on the blue colors ($u-g=-0.04$, $g-r=0.12$, $r-i=-0.10$, $i-z=0.03$). The spectrum shows double-peaked Balmer emission lines characteristic for eclipsing systems. Time series over two nights 27 days apart have shown the deep asymmetric eclipses from $g$=18.2 to 20.8 with the rather uncertain period of 78.12±0.45 minutes. Eclipses last for 3 minutes, starting with a fast drop from $g$=18.4 to 20.0 followed by slow fade to 20.8 and the abrupt rise to 18.6 (see Fig. 4 in Szkody et al.). Littlefair et al., 2008 have measured the orbital period to be 78.106657 minutes and concluded the system to be the period bouncer. Littlefair et al., 2013 have detected the donor in the $J$-band, establishing its sub-stellar nature and spectral type L2±1 (brown dwarf). Hernandez Santisteban et al., 2016 reported L1±1 type of the secondary changing to M9-L0 at phase 0.5 (on the dayside of the donor). They conclude that the brown dwarf in J1433+1011 has undergone the transition from its originally stellar to the sub-stellar regime due to the mass loss. Savoury et al., 2018 have determined the system parameters from eclipse modelling. The system has mass ratio $q$=0.0661±0.0007, white dwarf mass 0.865±0.005 $M_\odot$ and the donor mass 0.0571±0.0007 $M_\odot$. Gaia early data release 3 (Gaia collaboration, 2020) provides the following parallax and proper motion: Plx=4.286±0.157 mas, PM=54.66±0.27 mas/yr, corresponding to the distance modulus (m-M) of 6.8 and tangential velocity 60 km/s.

Thorstensen, 2020 has listed J1433+1011 among the non-outbursting objects, informally naming them "*lurkers*". According to his study, 22 objects out of 179 non-magnetic SDSS-selected CVs have never shown outbursts, despite being otherwise similar to quiescent dwarf novae. 10 out of 22 CVs without known outbursts were labeled by Thorstensen as DN-W (objects showing a white dwarf contribution in spectra). These are objects with low disc luminosities, low accretion rates and short orbital periods. For 7 of 10 objects the orbital periods have been measured, six of them being shorter than 100 minutes (from 78 to 99 minutes), with J1433+1011 having the shortest one.

## 2  Outburst detection

The outburst of J1433+1011 was initially suspected by the author on March 22, 2021 while checking Zwicky Transient Facility (Bellm et al., 2019) data on the variables from Table 5 in

Thorstensen, 2020. ZTF Lasair light curve of J1433+1011 updated in the real time is available at https://lasair.roe.ac.uk/object/ZTF18abcejeg with the snapshots of individual detections. Figure 1 shows ZTF light curve as of March 26, 2021 with the apparent fading trend after the winter conjunction with the Sun.

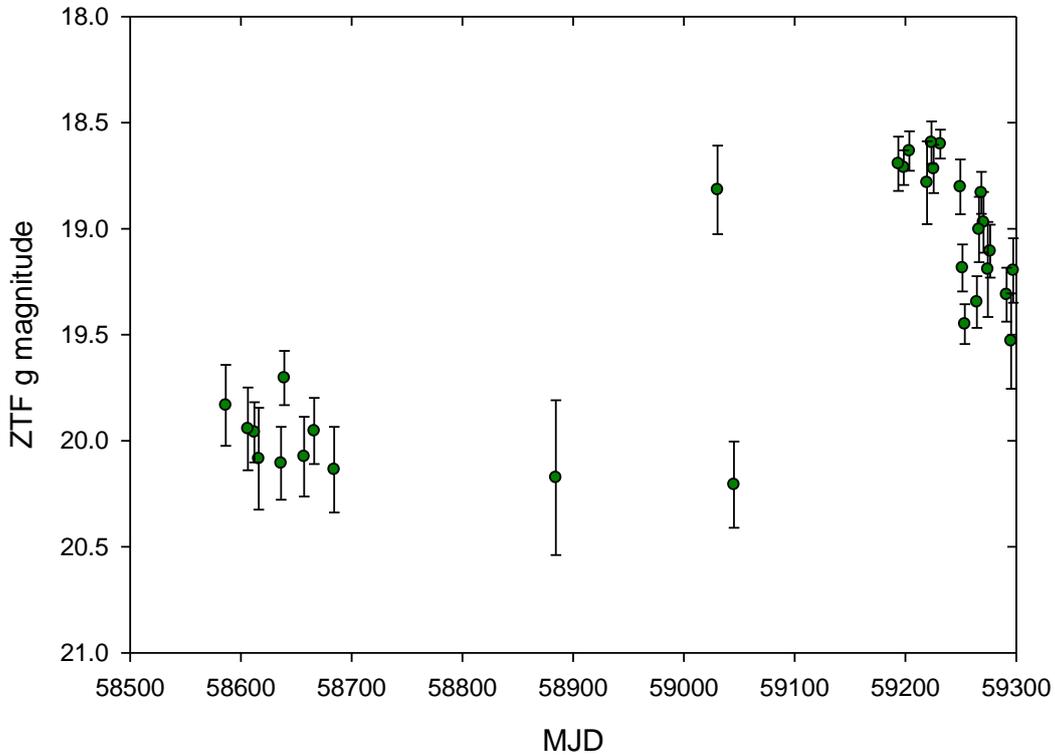

**Figure 1.** Light curve of SDSS J143317.78+101123.3 from ZTF Lasair data. MJD=59300 is 2021 Mar. 27.00 UT.

Since ZTF data looked like the tail of the long outburst (most likely the superoutburst), it caused the idea to check other publicly available surveys and databases. The recently released ATLAS forced photometry server (Tonry et al., 2018; Heinze et al., 2018) has confirmed the outburst. Moreover, it has provided the date of its start with the uncertainty of 5 days, between 2020 Sep. 05 and Sep. 11. The object was still at quiescence (18.6-18.9 magnitude) on four images obtained in $o$ (orange) filter between Sep. 05.231-05.252 UT, but definitely in an outburst to 13.67-13.36 magnitude in $c$ (cyan) filter on Sep. 11.225-11.250 UT. It has slightly faded to 13.83-14.04$o$ two nights later, on Sep. 13.220-13.240 UT. Next observations by ATLAS are dated by Dec. 06 when the star has faded by 4 magnitudes, to 17.9-18.3$o$. Even by the end of March 2021 the variable has not quite yet returned to its quiescence level.

Figure 2 shows the complete light curve of SDSS J1433+1011 from ATLAS data in 6 years since July 2015 to March 2021 with the detailed outburst light curve displayed in Figure 3. Though the images are not available from ATLAS database, the reality of the September outburst leaves no doubt. The errors of photometric measurements are 0.003-0.005$^m$ in $c$ filter on Sep. 11 and 0.007-0.016$^m$ in $o$ filter on Sep. 13 with the 5-σ limiting magnitude of 18-19 and 17.5 on two nights.

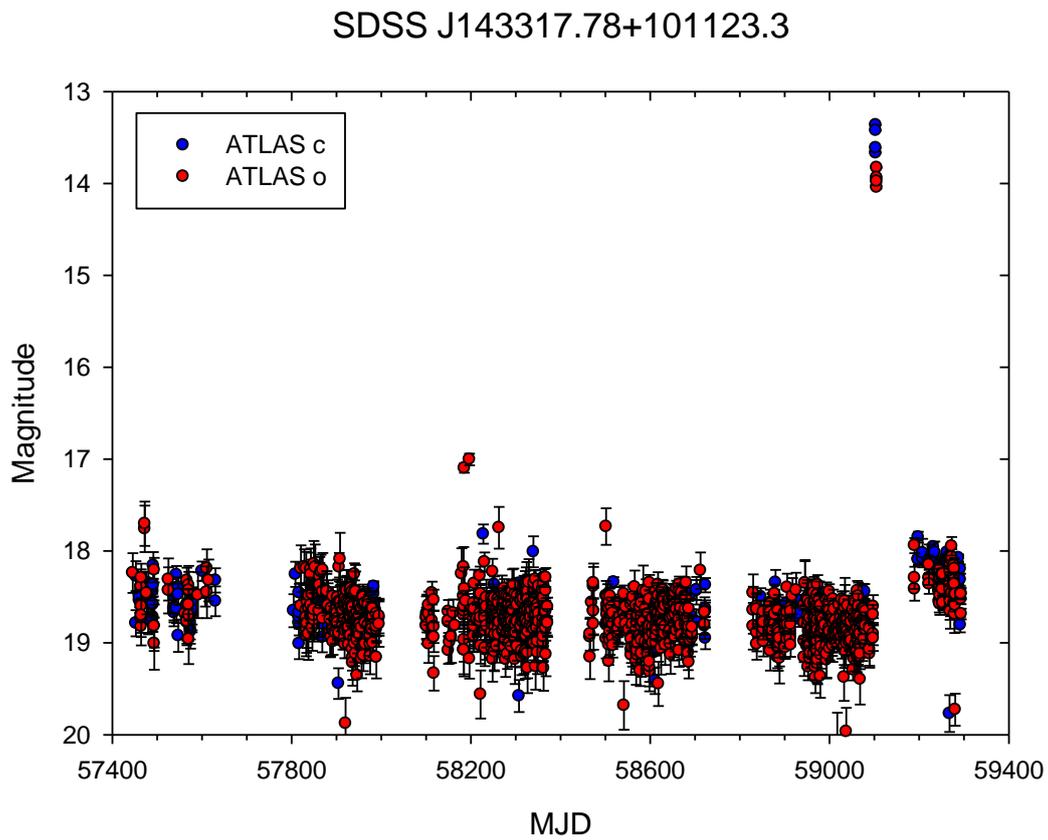

**Figure 2.** Light curve of SDSS J143317.78+101123.3 from ATLAS data.

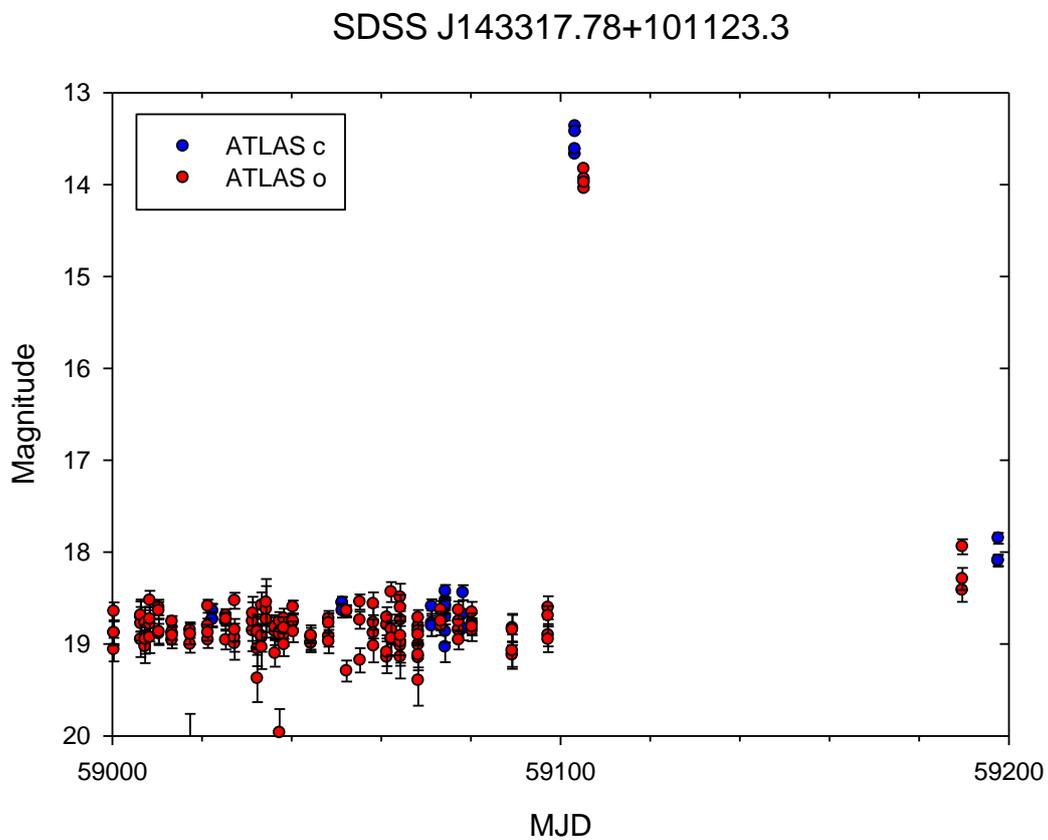

**Figure 3.** Zoomed light curve of 2020 outburst from ATLAS data. MJD=59100 is 2020 Sep. 08.00 UT.

Checking ASAS-SN database (Shappee et al., 2014; Kochanek et al., 2017) shows the last pre-outburst observation on Sep. 04.72 UT with the rather shallow upper limit of 15.7 magnitude in *g* (green) filter caused by the nearly full Moon with phase 0.94.

Table 1 provides the observing log of J1433+1011 in 2020-2021 with the last non-detection by surveys before the start of outburst, first observation after the seasonal break and the corresponding magnitudes or upper limits in the indicated photometric bands. *C* stands for Clear (unfiltered), not to be confused with *c* (cyan).

| Survey or observer name | Last observation | Magnitude or upper limit | First observation | Magnitude or upper limit |
|---|---|---|---|---|
| Eddy Muyllaert | 2020-06-25.97 | <14.3*V* | 2021-01-26.21 | <17.9*C* |
| ZTF | 2020-07-15.23 | 18.3*g* | 2020-12-01.55 | 18.0*r* |
| ASAS-SN | 2020-09-04.72 | <15.7*g* | 2020-12-02.52 | <15.8*g* |
| ATLAS | 2020-09-05.25 | 18.7*o* | 2020-12-06.63 | 17.9*o* |

**Table 1.** Observation details of J1433+1011 in 2020-2021 before and after the superoutburst.
Sources: https://www.aavso.org/LCGv2/ (E. Muyllaert, AAVSO); https://lasair.roe.ac.uk/conesearch/ (ZTF Lasair); https://asas-sn.osu.edu/ (ASAS-SN Sky Patrol); https://fallingstar-data.com/forcedphot/ (ATLAS).

We can see that ASAS-SN has missed the outburst by less than 6 days, while ZTF had stopped observing the area of interest almost two months before the outburst.

Formally speaking, the 2020 outburst cannot be declared a superoutburst since the superhumps were not observed by the obvious reasons. But the amplitude and duration of the outburst along with the short orbital period leave a little doubt regarding the UGWZ nature of J1433+1011. That is why we can take courage to call it a superoutburst.

The fading rate in ZTF *g*-filter data in December 2020 – March 2021 is 0.0072±0.0014 m/day, or 1 mag/140 days. This can be compared to the light curves of WZ Sge-type dwarf novae outbursts described in Kato, 2015. However, it should be made with caution since the light curves in 2015 paper are given for the time intervals 50-70 days long. Type D superoutburst of GW Lib shown in Fig. 8 of Kato, 2015, clearly demonstrates the fading rate slowing down over 30 days of the post-plato phase.

During the ATLAS detection on 2020 Sep. 11 the object was at the solar elongation of 48.5 degree (altitude 29° at the end of astronomical twilight). This is 9 degrees further from the Sun than the first asteroid inside the orbit of Venus 2020 AV$_2$ was at the moment of its discovery by ZTF (Ip et al., 2020). The first interstellar comet 2I/Borisov was discovered at the elongation of 38 degree. While it is obvious that the photometric time series longer than 45 minutes would not be possible at elongations that short, even the fact of outburst detection would be extremely valuable for such high-priority object. If it were detected and announced on short notice, it would have allowed for the outburst profile and initial fading rate to be determined by the nightly snapshots.

## 3 Discussion

Quoting Thorstensen, 2020, "It would not be surprising if all of DN-W systems in Table 5 were to outburst within the next couple of decades, but it remains possible that their outburst intervals could stretch still longer". Those words were written in May 2020 (article was submitted to arXiv on May 5). J1433+1011 has responded to them in September. However, being the "lurker", it has escaped the detection at the date of outburst. This study has brought it to light, confirming the brilliant prediction by Thorstensen.

SDSS J143317.78+101123.3 is located 20' north-north-east of the spiral galaxy NGC 5669, host of supernova SN 2013ab. During its explosion in 2013 this supernova was observed since Feb. 17 to Sep. 05 by many observers, according to the Recent Supernovae page by David Bishop. This shows that the outburst of J1433+1011 was potentially within the reach of supernova hunters who tend to observe galaxies with the past SN explosions. Checking the past images of the field of NGC 5669 is encouraged for the possible detections of J1433+1011.

Another possible source of the J1433+1011 outburst detections is in the field of comet observers. Namely, in November 2020 comet C/2020 P1 (NEOWISE) was passing 1.1° south of NGC 5669 and 1.5° south of J1433+1011. Minor Planet Center database has observations of this comet on Nov. 11 ($15.8^m$) and Nov. 29, but not on Nov. 20 when the comet was passing at minimum distance from our elusive object. Gennady Borisov has covered the southern part of Bootes searching for comets in the morning twilight of 2020 Nov. 27 with his 2.2x2.2° wide field instrument but has missed NGC 5669 and J1433+1011 by 0.7° (Borisov, private communication, 2021 Mar. 29).

The outbursts of low-accretion rate cataclysmic variables, especially those of period bouncers, are extremely rare, being literally once-in-a-lifetime events. We can have to wait for another outburst of J1433+1011 for 50 years or more. The ever-growing volume of observations by large area surveys and amateur astronomers can be more effective, with the special attention paid to the objects like those listed in Table 5 of Thorstensen, 2020.

Among the objects to be added to this watching list is the variable DDE 182 = IPHAS J192530.54+155426.5 (see http://scan.sai.msu.ru/VarDDE.html for the details) with the orbital period of $0^d.0539022$, strong Hα emission line and the shape of the phased light curve remaining stable over six years. Unlike J1433+1011, it does not show the eclipses and has a larger absolute magnitude, implying the smaller inclination angle. The parameters of six short-period "lurkers" among SDSS CVs and DDE 182 are given in Table 2. Objects are listed in order of orbital period.

| Object name | Coordinates | Mag | Distance, pc | $M_{abs}$ | Period, min |
|---|---|---|---|---|---|
| DDE 182 | 19 25 30.55 +15 54 26.6 | 18.07 | 500±29 | 9.6 | 77.62 |
| J1433+1011* | 14 33 17.78 +10 11 23.3 | 18.55 | 233±9 | 11.7 | 78.11 |
| V1247 Her | 17 11 45.08 +30 13 20.0 | 20.16 | 425±75 | 12.5 | 80.35 |
| J0043-0037 | 00 43 35.14 -00 37 29.8 | 19.85 | 411±80 | 11.8 | 83.39 |
| J0904+0355* | 09 04 03.48 +03 55 01.2 | 19.31 | 267±49 | 12.2 | 86.00 |
| J0039+0054 | 00 39 41.06 +00 54 27.5 | 20.76 | 1080±420 | 10.6: | 91.39 |
| J1216+0520 | 12 16 07.03 +05 20 13.9 | 20.11 | 354±87 | 12.4 | 98.83 |

* eclipsing system

**Table 2.** List of short-period objects (dwarf nova candidates) in need of continuous monitoring. Mag is *G* magnitude from Gaia EDR3. It may slightly differ from *G* mag from Gaia DR2 as listed by Thorstensen, 2020.

**Acknowledgments:** This work has used ATLAS forced photometry server. The Asteroid Terrestrial-impact Last Alert System (ATLAS) project is funded by NASA grants NN12AR55G, 80NSSC18K0284, and 80NSSC18K1575. ATLAS project is also partially funded by Kepler/K2 grant J1944/80NSSC19K0112 and HST GO-15889, and STFC grants ST/T000198/1 and ST/S006109/1. We acknowledge with thanks the observations of SDSS J1433+1011 from the AAVSO International Database contributed by Eddy Muyllaert, Belgium (observer code MUY).